\documentstyle[aps,preprint]{revtex}
\begin{document}

\title{Stable Quantum Monte Carlo Algorithm for $T=0$ 
Calculation  of Imaginary Time Green Functions}
\author{ F.F. Assaad and    M. Imada \\
         Institute for Solid State Physics, University of Tokyo,  \\
        7-22-1 Roppongi,
         Minato-ku, Tokyo 106, Japan. }

\maketitle

\begin{abstract}
We present a numerically stable Quantum Monte Carlo algorithm to
calculate zero-temperature 
imaginary-time Green functions $ G(\vec{r}, \tau) $
for Hubbard type models.
We illustrate the efficiency of the algorithm by calculating
the on-site Green function
$ G(\vec{r}=0, \tau) $ on $4 \times 4$ to $12 \times 12$ lattices
for the two-dimensional half-filled repulsive
Hubbard model at $U/t = 4$. 
By fitting the tail of $ G(\vec{r}=0, \tau) $ at long imaginary time
to the form
$e^{-\tau \Delta_c}$, we obtain a precise estimate
of the charge gap: $\Delta_c = 0.67 \pm 0.02$ in units of the hopping 
matrix element. We argue that the algorithm 
provides a powerful tool to study the metal-insulator
transition from the insulator side. 
\end{abstract}
\vspace{2cm}
Submitted to J. Phys. Soc. Jpn.

\newpage
\section{ Introduction}

The purpose of this article is to describe a numerically  stable 
Quantum Monte Carlo (QMC) algorithm to calculate zero-temperature 
imaginary time displaced Green functions:
\begin{eqnarray*}
G_{x,y}(\tau)  =  \Theta(\tau)
    \frac{ \langle \Psi_0 | c_{x}(\tau) c_{y}^{\dagger} |  \Psi_0 \rangle} 
           {\langle \Psi_0 |  \Psi_0 \rangle}
               - \Theta(-\tau)
    \frac{ \langle \Psi_0 | c_{y}^{\dagger}(-\tau) c_{x} |  \Psi_0 \rangle} 
           {\langle \Psi_0 |  \Psi_0 \rangle },
\end{eqnarray*}
\begin{equation}
            {\rm where }  \; \; \; \; 
            c_{x}(\tau)  =  e^{ \tau (H - \mu N) } c_{x}
                            e^{-\tau (H - \mu N)}.
\end{equation}
Here $ |\Psi_0 \rangle $ denotes the ground state of the
considered Hamiltonian $H$, 
$c_{x}^{\dagger}$ creates an electron with quantum numbers $x$, 
$\Theta(\tau)$ is the Heaviside function and
$\mu $ is the chemical potential which has to satisfy:
\begin{equation}
\label{chem}
	\lim_{\beta \rightarrow \infty} 
       \frac{ {\rm Tr} \left( e^{-\beta (H - \mu N)} N \right) }
            {  {\rm Tr} \left( e^{-\beta (H - \mu N)} \right) }
   =
        \frac{ \langle \Psi_0 | N |  \Psi_0  \rangle }
           {\langle \Psi_0 |  \Psi_0 \rangle },
\end{equation}
in the zero-temperature limit $T \equiv 1/\beta \rightarrow 0$.
The above equation implies that the chemical potential corresponding
to the desired particle density, $n(\mu)$, has to be 
known prior to the simulation. 
In a metallic state $n(\mu)$ is in general not known a priori.
However, in an insulating state at zero-temperature $n(\mu)$
is constant and  
in general known for chemical potentials within the charge gap $\Delta_c$.
In this situation, the here described algorithm proves to be a
powerful tool. 
The above $T=0$ Green functions  have already been calculated with QMC methods
by Deisz et al \cite{John}. Since their algorithm does not incorporate
a numerical stabilization scheme, they are restricted
to relatively small values of $\tau$ 
(i.e. $\tau  = 2.5 $ in units of the hopping 
matrix element for the one-dimensional Hubbard model).
This articles follows the work of Deisz et al. and describes
a numerical 
stabilization scheme which allows one to calculate $T=0$ Green functions
for arbitrary values of $\tau$. 

To demonstrate the efficiency of the algorithm, we calculate 
$G_{x,y}(\tau) $ for the two-dimensional half-filled Hubbard
model:
\begin{equation}
\label{Hubb}
H  =   \sum_{\vec{i},\vec{j}, \sigma}
       c_{\vec{i},\sigma}^{\dagger}T_{\vec{i},\vec{j}} c_{\vec{j},\sigma}
      + U \sum_{\vec{i}} \left( n_{\vec{i}, \uparrow} - \frac{1}{2} \right)
                         \left( n_{\vec{i}, \downarrow} - \frac{1}{2} \right).
\end{equation}
The quantum numbers $\vec{i}$ and  $\sigma$  denote lattice site and
{\it z}-component of spin respectively, 
$n_{\vec{i},\sigma } =  c_{\vec{i},\sigma}^{\dagger}
c_{\vec{i}\sigma}$ , and $T_{\vec{i},\vec{j}} = -t$ if
$\vec{i}$ and $\vec{j}$ are nearest-neighbors. In this
notation half-band filling corresponds to $\mu = 0$.
 As a non-trivial test of the  algorithm, one may fit the tail of 
$G_{x,x}(\tau)$ ($x = (\vec{i}, \sigma))$ 
 to the form $ e^{-\tau \Delta_c}$ to obtain
the charge gap $\Delta_c$.  At $U/t = 4$ and after 
extrapolation to the thermodynamic limit, our QMC data yields
$\Delta_c/t = 0.67 \pm 0.02 $. This value stands in good agreement
with previously determined values of $\Delta_c$
\cite{Furukawa}.

The article is organized in the following way. In the next section,
we briefly describe the zero-temperature auxiliary-field QMC
algorithm for the Hubbard model \cite{Koonin,Sandro,Imada,Assaad}.
We then present our solution for
the numerical stabilization of the time displaced Green
functions.  In section 3 we describe our
calculation of the charge gap for the two-dimensional Hubbard
model at $U/t = 4$. 
In the last section, we draw some conclusions and discuss the
potential applications of the algorithm. 

\section{The zero-temperature QMC algorithm }

Since the Hubbard model conserves particle number, and we are
working in the canonical ensemble, one may factorize the chemical
potential to obtain:
\begin{equation}
\label{factor}
G_{x,y}(\tau)  =  \Theta(\tau)
  \left.  \frac{ \langle \Psi_0 | c_{x}(\tau) c_{y}^{\dagger} |  \Psi_0 \rangle }
           {\langle \Psi_0 |  \Psi_0 \rangle } \right|_{\mu = 0} e^{\tau \mu}
               - \Theta(-\tau)
  \left.  \frac{ \langle \Psi_0 | c_{y}^{\dagger}(-\tau) c_{x} |  \Psi_0 \rangle }
           {\langle \Psi_0 |  \Psi_0 \rangle }\right|_{\mu = 0} e^{\tau  \mu}.
\end{equation}
Due to the above relation we consider the calculation of the 
$T=0$ Green functions at $\mu = 0$. 

The idea behind the zero temperature QMC algorithm is to filter
out the ground state from a trial wave function $| \Psi_T\rangle $ which is
required to be non-orthogonal to the ground state:
\begin{equation}
\frac{\langle \Psi_0 | c_{x}(\tau) c_{y}^{\dagger} |  \Psi_0 \rangle }
           {\langle \Psi_0 |  \Psi_0 \rangle }
           = \lim_{ \Theta \rightarrow \infty }
  \frac{ \langle \Psi_T |e^{-\Theta H }
          c_{x}(\tau) c_{y}^{\dagger}
         e^{-\Theta H } | \Psi_T \rangle }
       { \langle \Psi_T |e^{-2\Theta H } | \Psi_T \rangle }, \; \; \tau > 0.
\end{equation}

The QMC calculation of
\begin{equation}
 G^{>}_{x,y}(\Theta,\tau) \equiv \frac{ \langle \Psi_T |e^{-\Theta H }
          c_{x}(\tau) c_{y}^{\dagger}
         e^{-\Theta H } | \Psi_T \rangle }
       { \langle \Psi_T |e^{-2\Theta H } | \Psi_T \rangle }
\end{equation}
proceeds in the following way. 
The first step is to carry out a Trotter 
decomposition of the imaginary time propagation:
\begin{equation}
\label{Trotter}
      e^{-2\Theta H  }  =
 \left( e^{- \Delta \tau H_t/2 } e^{- \Delta \tau H_U} 
        e^{- \Delta \tau H_t/2 } \right)^{m} +  O( (\Delta \tau)^2).
\end{equation}
Here, $H_t$ ($H_U$) denotes the kinetic (potential) 
term of the Hubbard model and $m \Delta \tau = 2\Theta$. 
Having isolated the two-body interaction term, $H_U$, one may carry
out a discrete Hubbard Stratonovitch (HS) 
transformation \cite{Hirsch} to obtain:
\begin{equation}
\label{HS}
      e^{-\Delta \tau H_U}  = C \sum_{\vec{s}}  \exp
      \left( \sum_{x,y} c_{x}^{\dagger} D_{x,y}(\vec{s}) c_{y} \right),
\end{equation}
where  $\vec{s}$ denotes a vector of HS Ising fields.
For the Hubbard model (\ref{Hubb}), we take:
\begin{equation}
D_{\vec{i}\sigma, \vec{j}\sigma'} (\vec{s}) =
\delta_{\sigma,\sigma'} \delta_{\vec{i},\vec{j}} 
{\rm cosh}^{-1} (\Delta \tau U/2) s_{\vec{i}} \sigma.
\end{equation}
The constant $C = {\rm exp} ( -\Delta \tau N  U/2)/ 2^N $ for the
$N$-site system will be dropped below. 
The imaginary time propagation may now be written as:
\begin{eqnarray}
 & &  e^{-2\Theta H  } =
    \sum_{\vec{s}} U_{\vec{s}}(2\Theta, 0) +  O( (\Delta \tau)^2) 
\nonumber \\  
    {\rm where} \; \; \;
  & &U_{\vec{s}}(2\Theta, 0) = \prod_{n = 1}^{m}  
e^{- \Delta \tau H_t/2 }
\exp \left(  \sum_{x,y} c_{x}^{\dagger} D_{x,y}(\vec{s}_n) c_{y} \right)
e^{- \Delta \tau H_t/2 }.
\end{eqnarray}
In the above notation, $ G^{>}_{x,y}(\Theta,\tau)$ is given by:
\begin{equation}
 G^{>}_{x,y}(\Theta,\tau) = \frac
{ \sum_{\vec{s}} \langle \Psi_T | U_{\vec{s}}(2\Theta, \Theta + \tau) 
         c_x U_{\vec{s}}(\Theta + \tau,\Theta)
        c_y^{\dagger} U_{\vec{s}}(\Theta,0) | \Psi_T \rangle }
{\sum_{\vec{s}} \langle \Psi_T | U_{\vec{s}}(2\Theta, 0) | \Psi_T \rangle }
+  O( (\Delta \tau)^2).
\end{equation}
The trial wave function is required to be a Slater determinant:
\begin{equation}
\label{Trial}
     |\Psi_T \rangle 
      = \prod_{n=1}^{N_p} \left( \sum_x c_{x}^{\dagger} P_{x,n} \right)
                 |0\rangle .
\end{equation}
Here $N_p$ denotes the number of particles and $P $ is an $N_s \times N_p$
rectangular matrix where $N_s$ is the number of single particle
states.
Since $U_{\vec{s}}(2\Theta, 0)$ describes the propagation of non-interacting
electrons in an external HS field, 
one may integrate out the fermionic degrees of freedom to obtain:
\begin{equation}
\label{G>}
 G^{>}_{x,y}(\Theta,\tau) = 
      \sum_{\vec{s}} P_{\vec{s}} 
         \left[ \left( B_{\vec{s}} (\Theta + \tau, \Theta)\right)
          G_{\vec{s}}(\Theta,\Theta) \right]_{x,y} + O( (\Delta \tau)^2).
\end{equation}
In the above equation,
\begin{eqnarray*}
    B_{\vec{s}} (\Theta_2, \Theta_1) = \prod_{n = n_1 + 1}^{n_2}
          e^{-\Delta \tau T /2} 
          e^{D(\vec{s_n})}
          e^{-\Delta \tau T /2} \; \; {\rm where } \; \;
n_1 \Delta \tau = \Theta_1 \; \; {\rm and} \; \;  n_2 \Delta \tau = \Theta_2,
\end{eqnarray*}
\begin{eqnarray*}
	M_{\vec{s}} = P^{T} B_{\vec{s}} (2\Theta, 0 ) P,
\end{eqnarray*}
\begin{eqnarray*}
	P_{\vec{s}} =  \frac{ \det(M_{\vec{s}}) }
      { \sum_{\vec{s}} \det(M_{\vec{s}}) } 
\end{eqnarray*}
\begin{eqnarray*}
{\rm  and} \; \; \;
         \left( G_{\vec{s}}(\Theta,\Theta) \right)_{x,y} =
      \frac { \langle \Psi_T | U_{\vec{s}}(2\Theta, \Theta) 
         c_x  c_y^{\dagger} U_{\vec{s}}(\Theta,0) | \Psi_T \rangle } 
          { \langle \Psi_T | U_{\vec{s}}(2\Theta, 0) | \Psi_T \rangle } = \\
	\left( I - B_{\vec{s}} (\Theta,0)P M_{\vec{s}}^{-1} 
       P^{T} B_{\vec{s}} (2\Theta,\Theta) \right)_{x,y}.
\end{eqnarray*}
Here $I$ is the unit matrix, $I_{x,y} = \delta_{x,y}$.
In the same notation one obtains:
\begin{eqnarray}
\label{G<}
 G^{<}_{x,y}(\Theta,\tau) 
     & \equiv & - \frac{ \langle \Psi_T |e^{-\Theta H }
          c_{y}^{\dagger}(\tau) c_{x}
         e^{-\Theta H } | \Psi_T \rangle }
       { \langle \Psi_T |e^{-2\Theta H } | \Psi_T \rangle } \nonumber \\
     & = & 
         \sum_{\vec{s}} P_{\vec{s}} \left[
    \left( G_{\vec{s}}(\Theta,\Theta) - I \right)
     B_{\vec{s}}^{-1}(\Theta + \tau, \Theta) \right]_{x,y},  \; \; \tau > 0.
\end{eqnarray}
Summarizing, 
the zero-temperature imaginary-time Green function may be calculated from:
\begin{equation}
G_{x,y}(\tau)  =  \lim_{\Theta \rightarrow \infty }  \left( 
          \Theta(\tau) G^{>}_{x,y}(\Theta,\tau)  + 
          \Theta(-\tau) G^{<}_{x,y}(\Theta,-\tau)  \right) 
         + O( (\Delta \tau)^2).
\end{equation}

At half-band filling and due to particle hole symmetry, one may chose
a trial wave function such that $P_{\vec{s}}$ is positive definite.
$P_{\vec{s}} $ may be interpreted as a probability distribution 
and sampled with Monte-Carlo methods. 

\subsection{ Numerical Stabilization}
	The origin of the numerical instabilities occurring in the calculation
of Green functions may be understood by considering free electrons on a
two-dimensional square lattice. 
\begin{equation}
H =  -t \sum_{<\vec{i},\vec{j}>}
       c_{\vec{i}}^{\dagger} c_{\vec{j}}.
\end{equation}
Here, the sum runs over nearest-neighbors. For this Hamiltonian one has:
\begin{equation}
    \langle \Psi_0 | c_{\vec{k}}^{\dagger}(\tau) c_{\vec{k}} | \Psi_0 \rangle =
\exp \left(  \tau (\epsilon_{\vec{k}} - \mu)   \right) 
    \langle \Psi_0 | c_{\vec{k}}^{\dagger} c_{\vec{k}} |  \Psi_0 \rangle,
\end{equation}
where $\epsilon_{\vec{k}} = -2t(\cos(\vec{k} \vec{a}_x) + 
\cos(\vec{k} \vec{a}_y) ) $, $\vec{a}_x$, $\vec{a}_y$ being
the lattice constants. 
The chemical potential satisfies equation (\ref{chem})  and we will assume
$ |  \Psi_0 \rangle $ to be non-degenerate. 
In a numerical calculation the eigenvalues and eigenvectors of the
above Hamiltonian will be known up to machine precision, $\epsilon$.
In the case $ \epsilon_{\vec{k}} - \mu > 0 $, 
$ \langle \Psi_0 | c_{\vec{k}}^{\dagger} c_{\vec{k}} |  \Psi_0 \rangle \equiv 0$.
 However,
on a finite precision machine the later quantity will take a value of
the order of $\epsilon$. When calculating
$ \langle \Psi_0 | c_{\vec{k}}^{\dagger}(\tau) c_{\vec{k}} |  \Psi_0 \rangle $ this
roundoff error will be blown up exponentially and the result for large
values of $\tau$ will be unreliable.
In order to circumvent this problem, one may do the calculation at finite
temperature and then take the limit of vanishingly small temperatures:
\begin{equation}
    \langle \Psi_0 | c_{\vec{k}}^{\dagger}(\tau) c_{\vec{k}} |  \Psi_0 \rangle =
     \lim_{\beta \rightarrow \infty} \frac
       { \exp \left(  \tau (\epsilon_{\vec{k}} - \mu)   \right) }
       { 1 +  \exp \left(  \beta (\epsilon_{\vec{k}} - \mu)   \right) }.
\end{equation}
Even if the eigenvalues are known only up to machine precision, the right hand
side of the above equation for large but finite values of $\beta$ is
a numerically stable operation. Although very simple, this example
reflects the underlying numerical instabilities occurring in the calculation
of the  Green functions. 

We now consider the calculation of 
\begin{eqnarray}
\label{GST}
  & & G_{\vec{s}} ( \Theta + \tau, \Theta) =
     B_{\vec{s}} (\Theta + \tau, \Theta) G_{\vec{s}} ( \Theta, \Theta) 
     \; \; {\rm and } \nonumber \\
  & & G_{\vec{s}} ( \Theta , \Theta + \tau) =
      \left( G_{\vec{s}} ( \Theta , \Theta)  - I \right)
     B_{\vec{s}}^{-1} (\Theta + \tau, \Theta)
\end{eqnarray}
required to compute $G^{>}_{x,y}(\Theta,\tau)$ (see equation (\ref{G>}))
and $G^{<}_{x,y}(\Theta,\tau)$ (see equation (\ref{G<})) respectively.
The equal-time Green functions, $G_{\vec{s}} ( \Theta, \Theta)$,
may be calculated to machine precision \cite{Sandro,Imada,Assaad}. 
The matrices $B_{\vec{s}} (\Theta + \tau, \Theta) $ contain scales
which grow and decrease exponentially with $\tau$. As in the above example,
a straightforward multiplication of both matrices will lead to numerical 
instabilities for large values of $\tau$. Here, the problem is much more
severe since the presence of the HS field mixes 
different scales. 
In order to circumvent this problem, we propose the
following stabilization scheme. 

Since the trial wave function is a Slater determinant, we can find a single
particle Hamiltonian, $H_0 = \sum_{x,y} c^{\dagger}_x (h_0)_{x,y} c_y $,
which has $ | \Psi_T \rangle $ as a non-degenerate ground state. 
The equal time Green functions may then be written as:
\begin{eqnarray}
\label{GS}
    \left( G_{\vec{s}} ( \Theta, \Theta) \right)_{x,y} \equiv & &
     \frac { \langle \Psi_T | U_{\vec{s}}(2\Theta, \Theta) 
            c_x  c_y^{\dagger} U_{\vec{s}}(\Theta,0) | \Psi_T \rangle }
         { \langle \Psi_T | U_{\vec{s}}(2\Theta, 0) | \Psi_T \rangle } =
     \nonumber \\
\lim_{\beta \rightarrow \infty }
     \frac { {\rm Tr} \left( e^{-\beta H_0} U_{\vec{s}}(2\Theta, \Theta) 
            c_x  c_y^{\dagger} U_{\vec{s}}(\Theta,0)  \right) }
    { {\rm Tr} \left( e^{-\beta H_0} U_{\vec{s}}(2\Theta, 0) \right) }  
   & &  = \lim_{\beta \rightarrow \infty } 
           \left( I + B_{\vec{s}} (\Theta,0)e^{-\beta h_0} 
                  B_{\vec{s}} (2 \Theta, \Theta) \right)^{-1}_{x,y}
\end{eqnarray}
The last equality follows after integration of the fermionic degrees of
freedom.  Inspiring ourselves from the work of Hirsch
\cite{Hirsch1} we calculate the  time displaced Green functions 
in equation (\ref{GST}) with:
\begin{eqnarray}
\lim_{\beta \rightarrow \infty }
     \left( 
	  \begin{array}{cc}
            I & B_{\vec{s}} (\Theta,0)e^{-\beta h_0} 
                B_{\vec{s}}(2 \Theta, \Theta + \tau )  \\
           -B_{\vec{s}}(\Theta + \tau, \Theta)   & I \\
          \end{array}
     \right)^{-1}  =  \nonumber \\
     \left( 
	  \begin{array}{cc}
	    G_{\vec{s}} ( \Theta, \Theta) &  
            G_{\vec{s}} ( \Theta, \Theta + \tau )  \\
	    G_{\vec{s}} ( \Theta + \tau , \Theta) &  
            G_{\vec{s}} ( \Theta + \tau, \Theta + \tau )  \\
          \end{array}
     \right)
\end{eqnarray}
For very large but finite values of $\beta$, we can calculate 
the left hand side of the above equation by using matrix
stabilization techniques developed for finite temperature QMC algorithms. 
The basic idea behind those numerical stabilization techniques
is to keep the different scales occurring in the matrices $B_{\vec{s}}$ 
separate 
(for a review see reference \cite{Loh}). This is achieved by decomposing
the matrices $B_{\vec{s}}$ into 
a $UDV$ form where $U$ is an orthogonal matrix, $D$ a 
diagonal matrix containing
the exponentially large and exponentially small scales, and $V$ a triangular
matrix. 
The calculation of the left hand side of the above equation is done 
in the following way:
\begin{eqnarray*}
     \left( 
	  \begin{array}{cc}
            I & B_{\vec{s}} (\Theta,0)e^{-\beta h_0} 
                B_{\vec{s}}(2 \Theta, \Theta + \tau )  \\
           -B_{\vec{s}}(\Theta + \tau, \Theta)   & I \\
          \end{array}
     \right)^{-1}  =  
     \left( 
	  \begin{array}{cc}
            I & U_1 D_1 V_1 \\
            U_2 D_2 V_2 & I \\
          \end{array}
     \right)^{-1}  =  
\end{eqnarray*}
\begin{eqnarray*}
     \left( 
	  \begin{array}{cc}
            V_2 & 0 \\
            0 & V_1 \\
          \end{array}
     \right)^{-1}  
     \left( 
	  \begin{array}{cc}
            \left(V_2 U_1 \right)^{-1} & D_1 \\
            D_2 &  \left(V_1 U_2 \right)^{-1} \\
          \end{array}
     \right)^{-1}  
     \left( 
	  \begin{array}{cc}
            U_1 & 0 \\
            0 &  U_2 \\
          \end{array}
     \right)^{-1}  = 
\end{eqnarray*}
\begin{eqnarray}
    \left( 
	  \begin{array}{cc}
            V_2 & 0\\
            0 & V_1\\
          \end{array}
     \right)^{-1}  
     \left( U_3 D_3 V_3\right)^{-1}
     \left( 
	  \begin{array}{cc}
            U_1 & 0 \\
            0 &  U_2 \\
          \end{array}
     \right)^{-1}.
\end{eqnarray}
In the above equation, the matrix $D_3$ contains only exponentially
large scales since the matrices 
$\left(V_2 U_1 \right)^{-1} $ and 
$\left(V_1 U_2 \right)^{-1} $ act as a cutoff to the exponentially
small scales in the matrices $D_2$ and $D_1$. Since the other 
matrices are all well conditioned, the final matrix multiplication 
is well defined. 

A convenient choice of $H_0$  is obtained in a basis where
the trial wave function may be written as:
\begin{equation}
    | \Psi_T \rangle = \prod _{n = 1}^{N_p}  \gamma_n^{\dagger} | 0 \rangle.
\end{equation}
In this basis, we define $H_0$ through
\begin{equation}
     H_0 \gamma_n^{\dagger} | 0 \rangle = 
     \left\{ 
	  \begin{array}{c}
             - \gamma_n^{\dagger} | 0 \rangle  \;\; {\rm if } \;\; 
                  \gamma_n^{\dagger} \gamma_n | \Psi_T \rangle = 
                | \Psi_T \rangle \\
             +  \gamma_n^{\dagger} | 0 \rangle  \;\; {\rm if } \;\;
                  \gamma_n^{\dagger} \gamma_n | \Psi_T \rangle = 0 \\
          \end{array}
     \right.
\end{equation}
(Here, the energy unit is set by the hopping matrix element $t$.)
For this choice of $H_0$ values of $\beta t \sim 40$ were well sufficient
to satisfy equation (\ref{GS}) within required numerical precision
\cite{Note1}.
The above numerical stabilization scheme was indeed successful in
all examined cases. 

\section{Evaluation of the charge gap for the two dimensional 
Hubbard model}

We carried out our simulations  
on $4 \times 4$ to $12 \times 12$ lattices
for the two-dimensional half-filled ($\mu = 0$) repulsive
Hubbard model (\ref{Hubb}) at $U/t = 4$. Periodic boundary
conditions were assumed. 
A spin singlet ground state of the kinetic energy in 
the Hubbard Hamiltonian was used as a trial wave function. 
We test the quality of this trial wave function on a
$6 \times 6$ lattice. 
Figure 1 plots 
$ \langle \Psi_T |e^{-\Theta H } O e^{-\Theta H } | \Psi_T \rangle /
      \langle \Psi_T |e^{-2\Theta H } | \Psi_T \rangle  $
as a function of the projection parameter $\Theta$ for 
$ O = S(\pi,\pi)/N  = \frac{4}{3N} \sum_{\vec{r}}
 \exp \left( i \vec{Q} \vec{r} \right)
 \vec{S}(\vec{r}) \cdot \vec{S}( \vec{0} )  $ (solid circles
 in Figure 1a) 
and 
$ O =  E/N = H/N - U/4$  (solid circles in Figure 1b).
Here $\vec{Q}= (\pi, \pi)/a$, 
$\vec{S}(\vec{r})$ is the spin operator on site $\vec{r}$ and
$N$ denotes the number of sites. 
Already for values of $\Theta t = 2.5$, both considered observables
have converged within our estimated statistical uncertainty. 
For comparison, we have plotted 
$ {\rm Tr } \left( e^{-2 \Theta H } O \right)  / 
{\rm Tr } \left( e^{-2\Theta H }  \right) $ for
the same observables (triangles in Figure 1).  Values of $\Theta t$ at
least twice as large are required to obtain approximate ground 
state results.  

 Another source of systematic errors comes from the discretization
of the imaginary time propagation.  In Table \ref{table1} the 
$\Delta \tau$ dependence of the energy and $S(\pi,\pi)$ is given. 
The data are obtained form the zero-temperature QMC algorithm
on a $6 \times 6$ lattice and at $2\Theta t = 5$.
The values at $\Delta \tau = 0$ are obtained from a least square fit
of the finite $\Delta \tau$ results to the form
$a + b (\Delta \tau )^2$. We carried out our simulations at
$\Delta \tau t = 0.125$. As may be seen from Table \ref{table1}, 
this value of $\Delta \tau t$ produces a systematic error contained
in the quoted errorbars for the energy and a systematic error
of less than $1 \% $  for $ S(\pi,\pi) $.

To obtain an estimate of the charge gap, we consider
\begin{equation}
 G(\vec{r} = 0, \tau) = \frac{1}{N} \sum_{x} G_{x,x}(\tau), 
\; \; \tau > 0.
\end{equation}
Here,  $x$ stands for spin and site indices. 
Inserting a complete set of eigenstates
of the Hamiltonian $H$ in the $N+1$ particle Hilbert space yields:
\begin{equation}
  G(\vec{r} = 0, \tau) = \frac{1}{N} \sum_{n,x} 
 | \langle \Psi_{0}^{N} | c_x | \Psi_{n}^{N +1} \rangle |^2 
   \exp \left( -\tau \left( E_n^{N+1} - E_0^N \right) \right).
\end{equation}
where $H | \Psi_{n}^{N +1} \rangle = E_n^{N+1} | \Psi_{n}^{N +1} \rangle $
and $H  | \Psi_{0}^{N} \rangle =  E_0^N  | \Psi_{0}^{N} \rangle $. 
At large values of $\tau t $, $G(\vec{r} = 0, \tau) \sim 
\exp \left( - \tau \Delta_c \right) $ where 
$ \Delta_c \equiv  E_0^{N+1} - E_0^N  $ corresponds to the charge gap.

Our results are plotted in Figure 2. For those simulations we
have chosen $\Theta t = 13.5$.  Since values of $\tau$ 
up to $\tau_{max} t =  12 $ were considered, the effective projection 
parameter is given by: $\Theta_{eff} = \Theta - \tau_{max}/2 = 7.5/t$.
As may be seen from Figure 1, this value of the projection parameter
is more than sufficient to filter out the ground state from the trial wave
function. 
The solid lines in Figure 2 correspond to least square fits of the tail
of $G(\vec{r} = 0, \tau)$ to the 
form $\exp( - \tau \Delta_c) $ \cite{Note2}. 
The estimated value of the gap as a function of system size is
plotted in Figure 3. A least square fit
of the data to the form $a + b/L$,  where $L$ denotes the linear
length of the lattice,  yields
$\Delta_c/t = 0.67 \pm 0.02$ in the thermodynamic limit.  
This value stands in good agreement with the value of the charge
gap obtained by Furukawa and Imada \cite{Furukawa}:
$\Delta_c/t = 0.58 \pm 0.08$
(solid circle in Figure 3.) As may be seen from the
comparison of errorbars, the accuracy of the estimation
has been much improved in the present study. 

\section{Conclusions}
We have presented an efficient, numerically stable, QMC algorithm
to calculate $T=0$ imaginary time Green functions for Hubbard type
models. As a non-trivial test application of this
algorithm, we have obtained an accurate
estimate of the charge gap for the two-dimensional
half-filled repulsive Hubbard model at $U/t = 4$:
$\Delta_c/t = 0.67 \pm 0.02$.

The algorithm is formulated in the canonical ensemble. Hence, the relation
$n(\mu)$ has to be known prior to the simulation. This
renders the algorithm hard to use in a metallic state.  However,
in an insulating state at zero temperature 
$n(\mu)$ is constant, and generally known,  
for chemical potentials within the
charge gap. In this
situation the here presented algorithm proves to be
a powerful tool. We illustrate this by considering the two-dimensional 
Hubbard model. Due to Equation (\ref{factor}), it suffices to know the
Green functions at $\mu = 0$ (half-filling) to be able to
determine  them trivially for any other chemical potential within
the charge gap. At $\mu = 0$, we are not confronted with a sign problem
due to particle-hole symmetry and the statistical  fluctuations
do not blow-up  exponentially with growing lattice sizes and projection
parameters $\Theta$.
It is however clear from equation (\ref{factor}) that statistical 
fluctuations will increase (decrease) exponentially with growing 
positive values
of $\tau$ for $\mu > 0$ ($\mu < 0$). 
In comparison, finite temperature algorithms in 
 the grand-canonical ensemble,
are faced with a sign problem away from $ \mu = 0$.  
Hence, statistical fluctuations grow exponentially with growing 
lattice size and inverse temperature. Away from $\mu = 0$, it is
thus extremely hard to extrapolate any zero temperature result 
from the finite temperature grand-canonical algorithms for large lattice
sizes. This renders the here presented algorithm a powerful
tool for the study of the metal-insulator transition from the insulator 
side \cite{Assaad1}. 

\section*{Acknowledgements}
F.F. Assaad thanks the JSPS for financial support.
The numerical calculations were carried out on the Fujitsu VPP500 of the 
Supercomputer Center of the Institute for Solid State Physics, Univ.
of Tokyo. This work is supported by a Grant-in-Aid for Scientific 
Research on the Priority Area "Anomalous Metallic State near the Mott
Transition" from the Ministry of Education, Science and Culture, Japan.

\subsubsection*{Figure captions}
\newcounter{bean}
\begin{list}%
{Fig. \arabic{bean}}{\usecounter{bean}
                   \setlength{\rightmargin}{\leftmargin}}

\item  $\bullet$: 
$ \langle \Psi_T |e^{-\Theta H } O e^{-\Theta H } | \Psi_T \rangle /
      \langle \Psi_T |e^{-2\Theta H } | \Psi_T \rangle  $
as a function of the projection parameter $\Theta$. The trial wave
function is a singlet ground state of the kinetic energy in
Hubbard Hamiltonian. 
$\triangle$: $ {\rm Tr } \left( e^{-2 \Theta H } O \right)  /
{\rm Tr } \left( e^{-2\Theta H }  \right) $.  \\
We have considered two observables: a)$O =  S(\pi,\pi)/N$, b) $ O = E/N$.

\item  $\ln G(\vec{r}=0, \tau) $ as a function of lattice size.
The solid lines are least square fits of the tail of $G(\vec{r}=0, \tau)$
to the form $e^{-\tau \Delta_c}$.

\item $\Delta_c$ as a function of $1/L$ where $L$ corresponds
to the linear size of the square lattice. The solid circle at 
$1/L = 0$ corresponds to Furukawa and Imada's estimate of $\Delta_c$
(see reference \cite{Furukawa}).
\end{list}

\begin{table}
\caption{ Energy per site and $S(\pi,\pi)$ as a  function of
$\Delta \tau t$ for the Hubbard model at $U/t=4$ on  a
$6 \times 6$ lattice.  The simulations were carried out with
the zero-temperature QMC algorithm at $2 \Theta t = 5$.
The quoted values at $\Delta \tau t = 0$ are obtained from a least
square fit to the form $a + b (\Delta \tau t)^2$ }
\label{table1}
\begin{tabular}{@{\hspace{\tabcolsep}\extracolsep{\fill}}ccc} \hline
$ \Delta \tau t$ & $ E/Nt $ & $ S(\pi,\pi)/N $ \\ \hline
$  0.0 $  & $-0.8575 \pm 0.0003$ & $0.1579 \pm 0.0007 $  \\
$\; 0.0625  \;$ &  $\;-0.8571 \pm 0.0004  \;$ &  $\; 0.1578 \pm 0.0009 \;$ \\
$0.1    $ &  $ -0.8571 \pm 0.0003 $ &  $ 0.1570 \pm 0.0006 $ \\
$0.125  $ &  $ -0.8570 \pm 0.0003 $ &  $ 0.1564 \pm 0.0007 $  \\
$0.166  $ &  $ -0.8563 \pm 0.0003 $ &  $ 0.1557 \pm 0.0008 $  \\
\hline
\end{tabular}
\end{table}

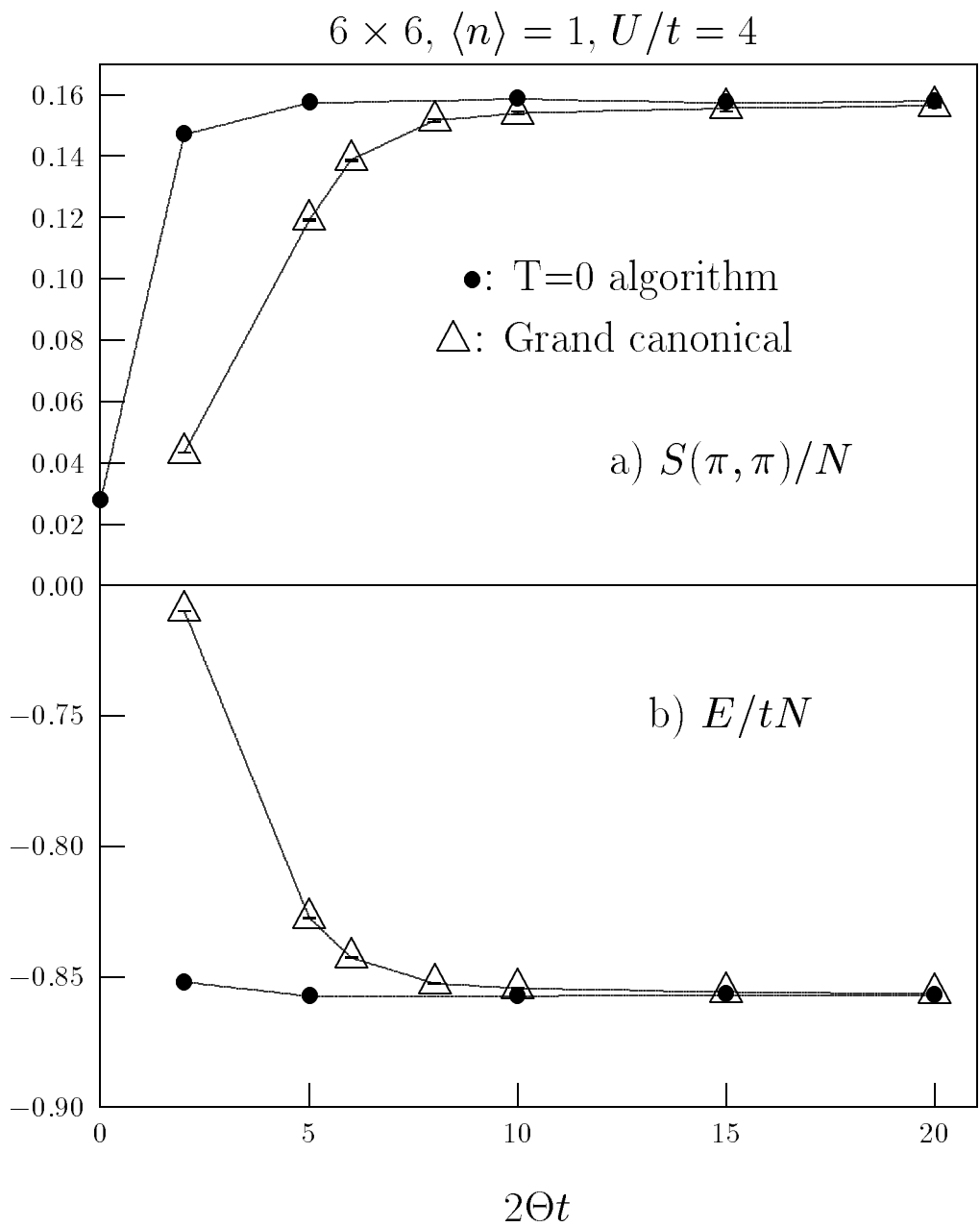
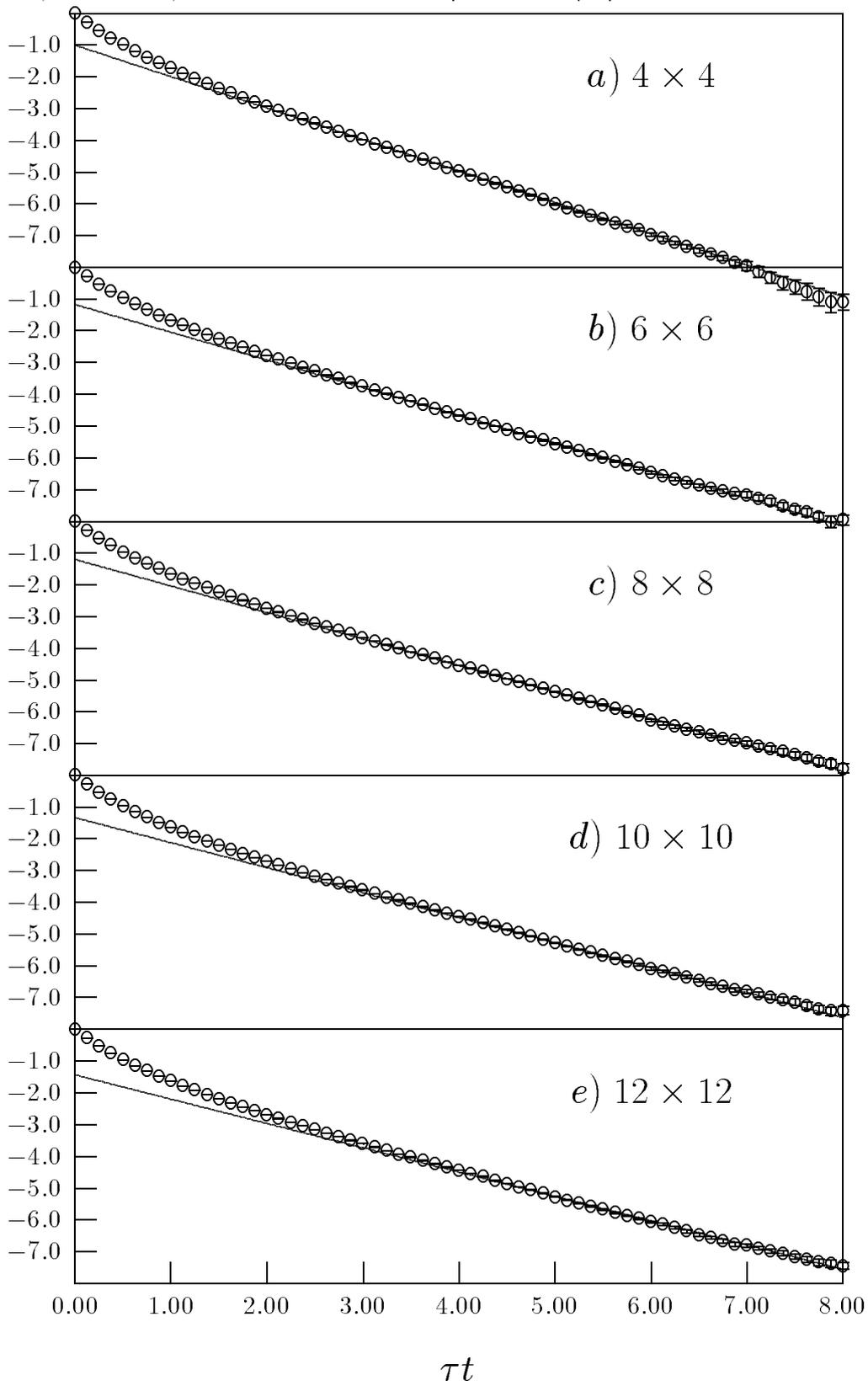
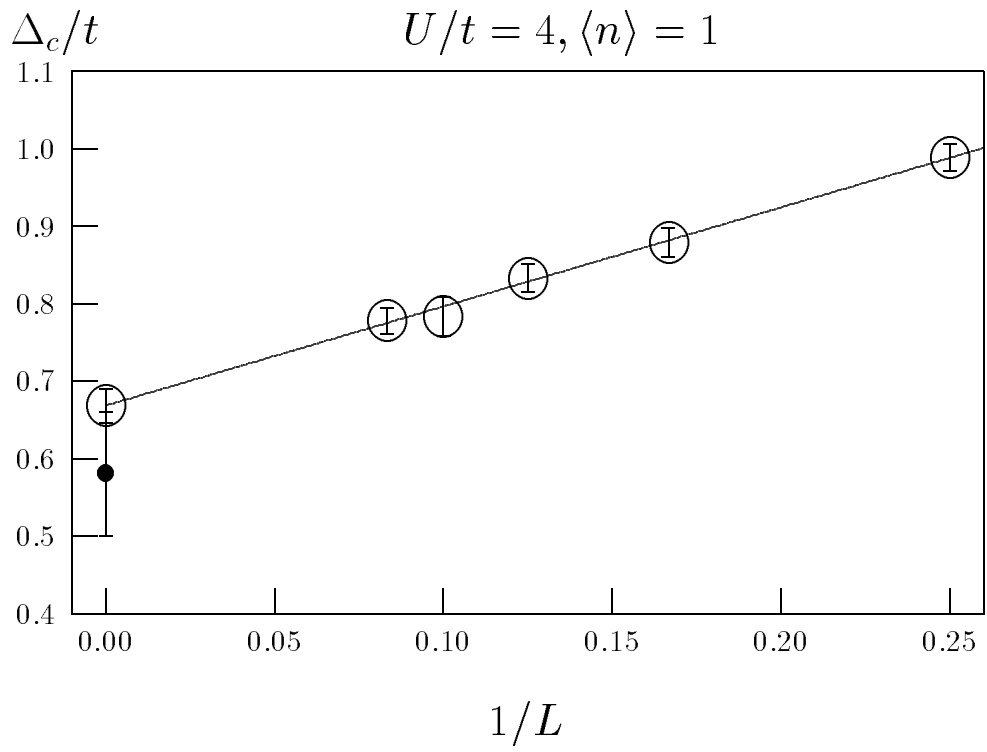
\end{document}